\documentclass[a4paper,fleqn,usenatbib]{mnras}
\usepackage{newtxtext,newtxmath}
\usepackage[T1]{fontenc}
\usepackage{ae,aecompl}
\usepackage{graphicx}	
\usepackage{amsmath}	
\usepackage{booktabs}	
\usepackage{setspace}
\usepackage{url}

\usepackage{CJKutf8}
\newcommand{\CNnames}[1]{{\begin{CJK}{UTF8}{gbsn}~(\textbf{#1})~\end{CJK}}}

\usepackage{tablefootnote,footnotehyper} 
\usepackage{array}
\newcolumntype{H}{>{\setbox0=\hbox\bgroup}c<{\egroup}@{}} 

\title[Pleiades $\delta$\,Sct stars]{Five young $\delta$\,Scuti stars in the Pleiades seen with Kepler/K2}

\author[Simon J. Murphy et al.]{
Simon J. Murphy,$^{1,2}$\thanks{E-mail: simon.murphy@sydney.edu.au (SJM)}
Timothy R. Bedding,$^{1,2}$\thanks{E-mail: tim.bedding@sydney.edu.au (TRB)}
Timothy R. White,$^{1,2}$\thanks{E-mail: tim.white@sydney.edu.au (TRW)} \and
Yaguang Li\CNnames{李亚光},$^{1,2}$
Daniel Hey,$^{1,2}$
Daniel Reese,$^{3}$ and
Meridith Joyce.$^{4}$
\\
$^{1}$ Sydney Institute for Astronomy (SIfA), School of Physics, University of Sydney, NSW 2006, Australia\\
$^{2}$ Stellar Astrophysics Centre, Department of Physics and Astronomy, Aarhus University, 8000 Aarhus C, Denmark\\
$^{3}$ LESIA, Observatoire de Paris, Universit\'e PSL, CNRS, Sorbonne Universit\'e, Universit\'e de Paris, 5 place Jules Janssen, 92195 Meudon, France\\
$^{4}$ Space Telescope Science Institute in Baltimore, MD, USA\\
}

\date{Accepted XXX. Received YYY; in original form ZZZ}
\pubyear{2021}
\begin{document}
\label{firstpage}
\pagerange{\pageref{firstpage}--\pageref{lastpage}}
\maketitle

\begin{abstract}
We perform mode identification for five $\delta$\,Scuti stars in the Pleiades star cluster, using custom light curves from K2 photometry. 
By creating \'echelle diagrams, we identify radial and dipole mode ridges, comprising a total of 28 radial and 16 dipole modes across the five stars. We also suggest possible identities for those modes that lie offset from the radial and dipole ridges. We calculate non-rotating stellar pulsation models to verify our mode identifications, finding good agreement within the age and metallicity constraints of the cluster. We also find that for all stars, the least dense models are preferred, reflecting the lower density of these oblate, rotating stars. Three of the five stars show rotationally-split multiplets. We conclude that the sample shows promise for asteroseismic rotation rates, masses, and ages with rotating models in the future. Our preliminary modelling also indicates some sensitivity to the helium abundance.
\end{abstract}

\begin{keywords}
asteroseismology -- stars: variables: $\delta$\,Scuti -- stars: evolution -- stars: oscillations -- star clusters: individual: Pleiades
\end{keywords}



\section{Introduction}
\label{sec:intro}

At 136\,pc and with $\sim$1200\,members, the Pleiades is a nearby and populous young open cluster that serves as a test-bed for stellar evolution \citep[e.g.][]{heyletal2021a,meingastetal2021}. In particular, clusters like the Pleiades are fundamental calibrators of secondary age indicators such as gyrochronology and lithium depletion, which were recently evaluated using late-type Pleiads \citep{bouvieretal2018} and Kepler/K2 rotation rates \citep{rebulletal2016a}.

It is therefore surprising that age spreads of up to 50\% for the Pleiades are found in the literature. For instance, non-rotating isochrones suggest an age as young as 100\,Myr \citep[e.g.][]{meynetetal1993}, while rotating isochrones suggest a wide range of plausible ages (100--160\,Myr; \citealt{brandt&huang2015b,gossageetal2018}). The lithium depletion boundary (LDB) suggests $130\pm20$\,Myr \citep{barradoetal2004,bouvieretal2018}, yet brown-dwarf members of the Pleiades suggest an LDB age of $112\pm5$ that is further reduced to $\sim$100\,Myr when magnetic effects are accounted for \citep{dahm2015}. Cluster ages appear sensitive to the choice of model physics \citep{choietal2016,gossageetal2018} and, in particular, to the treatment of stellar rotation \citep{brandt&huang2015a}.

Rotation has two important effects on the observed properties of stars. Firstly, it deforms stars to oblate spheroids through centrifugal distortion, leading to cooler gravity-darkened equators and brighter poles \citep{vonzeipel1924}. This causes the observed temperatures and luminosities to depend on both the rate and inclination angle of the rotation \citep{lucy1967,espinosa&rieutord2011}. Rapid rotators will be observed to be cooler and/or brighter (hence older) than an isochrone of their true age (Fig.\,\ref{fig:hrd}). Secondly, rotation drives large-scale mixing in stars \citep{endal&sofia1978,hegeretal2000,maeder&meynet2000}, which brings fresh hydrogen to the core and extends the main-sequence lifetime.
Both effects from rotation influence the main-sequence turn-off, explaining the wide range of isochrone ages in the literature.

Asteroseismology is one technique that can potentially overcome these problems. In particular, stellar rotation rates can be inferred from rotational splittings of the pulsation frequencies \citep{aertsetal2010}. When $v \sin i$ is available, the stellar inclination is then calculable under the assumption of solid-body rotation, and that assumption is testable \citep{dziembowski&goode1992,nielsenetal2015,benomaretal2018,vanreethetal2018,hattaetal2019}. Moreover, the stellar pulsation frequencies place strong constraints on the stellar density and internal structure \citep{rodriguez-martinetal2020,dhouibetal2021,michielsenetal2021}, including on angular momentum transport and chemical mixing \citep{mombargetal2019,mombargetal2021}, provided that the oscillation modes can be identified. Since the observational constraints can come directly from the seismology, the effects of gravity darkening on the observed temperature and luminosity become less problematic. In this context, the discovery that some young $\delta$\,Scuti stars oscillate in regular patterns that allow their modes to be identified has been a significant breakthrough \citep{beddingetal2020}.  This breakthrough was enabled by the ultra-precise and nearly-continuous light curves from the space telescopes {\it Kepler} \citep{boruckietal2010} and {\it TESS} \citep{rickeretal2015}. In this paper, we use {\it Kepler}/K2 observations to study young $\delta$\,Sct stars in the Pleiades open cluster.

The study of star clusters is undergoing a revolution of its own, thanks to ESA's {\it Gaia} spacecraft \citep[e.g.][]{cantat-gaudinetal2018,gaiacollaboration2018c,heyletal2021b,jacksonetal2021}. This raises the exciting prospect of asteroseismic ages for several clusters or associations. For instance, \citet{beddingetal2020} calculated an asteroseismic age of 150\,Myr for a $\delta$\,Sct star in the newly-discovered Pisces-Eridanus stream, whose age had been contentious \citep{curtisetal2019,meingastetal2019}. The same technique provided a precise age and metallicity for a pre-main-sequence $\delta$\,Sct star in the Upper Centaurus--Lupus subgroup of the Sco--Cen association \citep{murphyetal2021a}. The process of using asteroseismology to measure ages begins with the detection of pulsating stars and identification of their pulsation modes, which is what we do here for five $\delta$\,Sct members of the Pleiades. 

In 2014 the Kepler Mission was repurposed into K2 following the failure of a second of four angular momentum reaction wheels \citep{howelletal2014}. K2 observed a series of fields along the ecliptic, and Campaign~4 (72\,d duration) included more than 1000 stars in the Pleiades. Most of those stars were observed in long-cadence mode (30-min sampling), which leads to Nyquist ambiguity in the pulsation frequencies \citep{murphy2012a}. Nonetheless, those light curves revealed some $\delta$\,Sct pulsators \citep{rebulletal2016b}. Only six $\delta$\,Sct Pleiads are known from ground-based photometry \citep[][and references therein]{breger1972,fox-machadoetal2006}, and five of these were observed at short cadence (1-min sampling) by K2 (Sec.\,\ref{sec:obs}). We have created custom light curves for these five stars (Sec.\,\ref{sec:k2}), and used the Fourier amplitude spectra to identify their pulsation modes (Sec.\,\ref{sec:pulsations}).

We have also calculated preliminary pulsation models, described in Sec.\,\ref{sec:models}, using a narrow metallicity prior from the cluster metallicity. Clusters are believed to be chemically homogeneous \citep{desilvaetal2006,sestitoetal2007,bovy2016}, and this can be used to verify cluster members in a process known as chemical tagging \citep{kosetal2018}. Importantly, it means we can assume that all Pleaids have the same metallicity, which can be measured precisely using the narrow-lined members. \citet{soderblometal2009} analysed high-resolution and high-SNR spectra of 20 solar-type Pleiads, finding [Fe/H] $= +0.03 \pm 0.02$ (stat.) $\pm~0.05$ (sys.). We note that for A stars, and especially slow rotators ($v\sin i \lesssim 100$\,km\,s$^{-1}$), [Fe/H] tends to be above solar due to radiative levitation in poorly-mixed stellar atmospheres \citep{gebranetal2010,murphy2014}. The overall metal mass fractions, however, should agree with those of the late-type dwarfs, for which abundances can be measured reliably.

Early-type Pleaids are likely to be found in binary systems \citep{moe&distefano2017}, necessitating some caution in analyses of the cluster age \citep{gossageetal2018}. Binaries can be problematic when fitting isochrones because they cause stars to appear overly luminous in the Hertzsprung--Russell (H--R) diagram \citep{gaiacollaboration2018b}. \citet{torres2020} spectroscopically monitored 33 B and A-type Pleiads and found a lower limit to the binarity of 37\%, but only one of our five stars (V650\,Tau) was in that sample. The mass ratios of binaries with A-type primaries typically skew to low values, with no clear excess of twins \citep{murphyetal2018}. Hence, while we should expect that some stars will be binaries, it is unlikely that both components will show $\delta$\,Sct pulsations and so binarity should not affect our analysis.

With its lower susceptibility to rotation and to binarity, asteroseismology of $\delta$\,Sct stars using space-based photometry may eventually lead to luminosities that are more reliable than those calculated via parallaxes and visual magnitudes, particularly when extinction is large and/or uncertain. A period--luminosity relation for $\delta$\,Sct stars exists, albeit with considerable scatter \citep[e.g][]{breger&bregman1975,king1991,porettietal2008,mcnamara1997,mcnamara2011}. \citet{ziaalietal2019} and \citet{jayasingheetal2020} revisited the P--L relation for $\delta$\,Sct stars with Gaia DR2 parallaxes, tightening the relation. In the future, we expect asteroseismic models calculated with rotation to place tight constraints on individual stellar luminosities.

Our analysis of the five $\delta$\,Sct Pleaids begins with a review of observational constraints from the literature.

\section{The five $\delta$\,Sct stars}
\label{sec:obs}

\begin{figure}
\begin{center}
\includegraphics[width=0.48\textwidth]{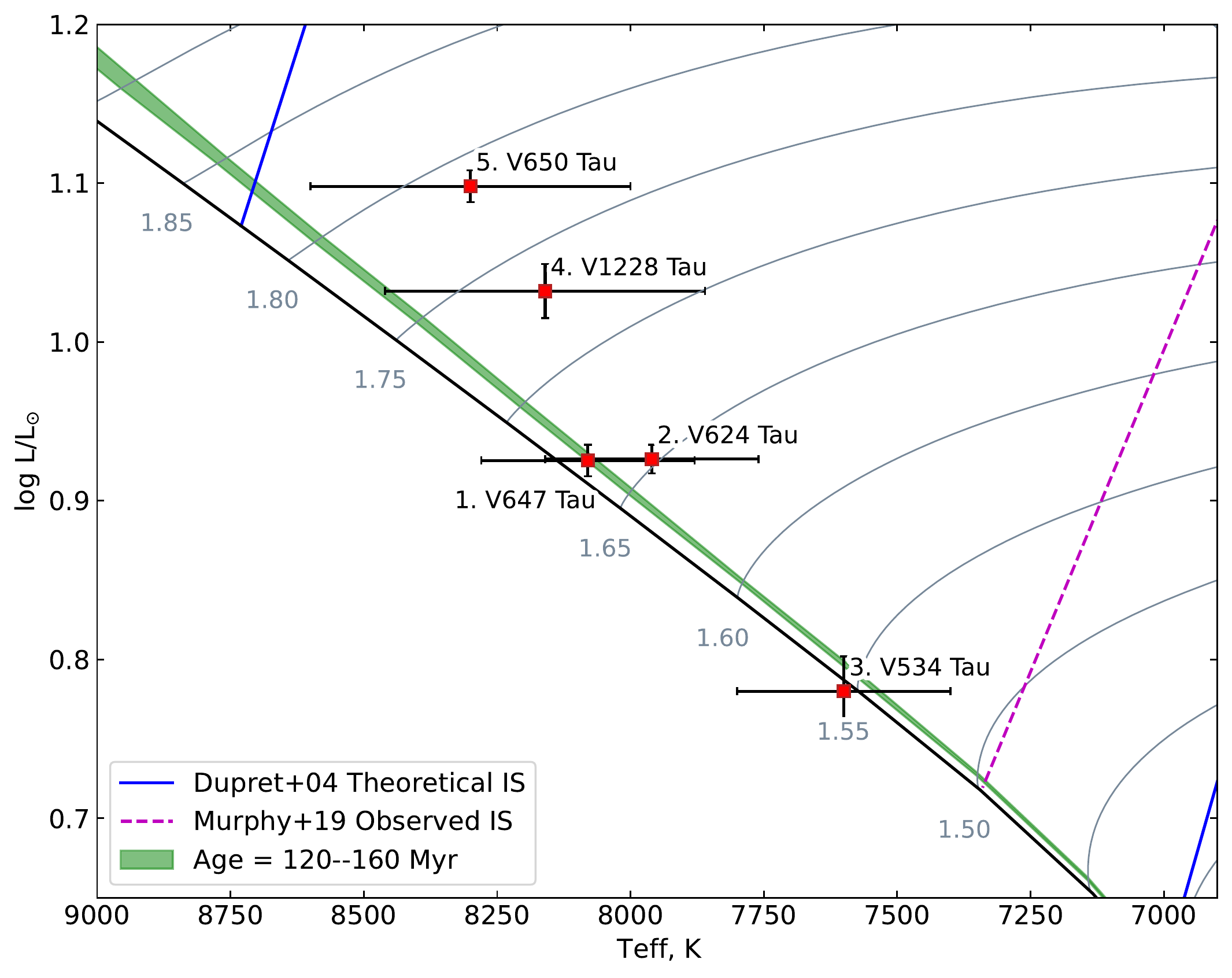}
\caption{H--R diagram showing the five $\delta$\,Sct stars relative to the cluster isochrone and the $\delta$\,Sct instability strip. Stars are numbered by increasing $v \sin i$. Evolutionary tracks of solar metallicity and the red edge of the observed instability strip edge are from \citet{murphyetal2019}; the theoretical instability strip is from \citet{dupretetal2004}. Numbers beneath the ZAMS (solid black line) show the mass of each track in solar masses.  The green strip shows the span of isochrones corresponding to the Pleaides age estimate of 120--160\,Myr.}
\label{fig:hrd}
\end{center}
\end{figure}

Table\:\ref{tab:identifiers} lists alternative identifiers, $V$ magnitudes and coordinates for the five stars, while Table\:\ref{tab:obs} summarises their properties, which are discussed in more detail in the following sections.  In these tables and throughout this paper, we order the stars by $v\sin i$.  The locations of the stars in the H--R diagram are shown in Fig.\,\ref{fig:hrd}.  We see that the two rapid rotators (V1228\,Tau and V650\,Tau) are displaced significantly from an isochrone of 120--160\,Myr created using the tracks shown (green band).

\begin{table*}
\centering
\caption{Alternative identifiers, $V$ magnitudes and coordinates of the five Pleiades $\delta$\,Sct stars discussed in this work.}
\label{tab:identifiers}
\begin{tabular}{rclccccc}
\toprule
GCVS & HD & \multicolumn{1}{c}{TYC} & EPIC & TIC & Vmag & RA (J2000) & Dec (J2000) \\
\midrule
V647\,Tau & HD\,23607 & TYC\,1800-1607-1 & EPIC\,211072836 & TIC\,405484188 & 8.26 & 03 47 19.34559 & +24 08 20.81410\\
V624\,Tau & HD\,23156 & TYC\,1799-73-1 & EPIC\,211088007 & TIC\,405483817 & 8.22 & 03 43 43.24180 & +24 22 28.44981 \\
V534\,Tau & HD\,23567 & TYC\,1804-2095-1 & EPIC\,211115721 & TIC\,405484574 & 8.30 & 03 47 03.57648 & +24 49 11.15476 \\
V1228\,Tau & HD\,23628 & TYC\,1804-1961-1 & EPIC\,211101694  & TIC\,125754823 & 7.71 & 03 47 24.07588 & +24 35 18.55427 \\
V650\,Tau & HD\,23643 & TYC\,1800-1630-1 & EPIC\,211044267 & TIC\,440681425 & 7.79 & 03 47 26.82832 & +23 40 41.98715 \\
\bottomrule
\end{tabular}
\end{table*}

\begin{table}
\centering
\caption{Adopted observed stellar properties and their uncertainties. Note that the cluster metallicity, \mbox{[Fe/H] = $+0.03\pm0.05$}, was used during modelling (Sec.\,\ref{sec:models}).}
\label{tab:obs}
\begin{tabular}{rccrr}
\toprule
\multicolumn{1}{c}{Name} & $T_{\rm eff}$ & \multicolumn{1}{c}{$\log L$/L$_{\odot}$} & \multicolumn{1}{c}{[Fe/H]} & \multicolumn{1}{c}{$v\sin i$} \\
\multicolumn{1}{c}{(GCVS)} & K & & \multicolumn{1}{c}{dex} & km\,s$^{-1}$ \\
\midrule
V647\,Tau & $8080\pm200$ & $0.928\pm0.019$ & $0.19\pm0.10$ & $10\pm3$ \\
V624\,Tau & $7960\pm200$ & $0.944\pm0.017$ &$0.17\pm0.10$ & $42\pm5$ \\
V534\,Tau & $7600\pm200$ & $0.780\pm0.022$ & \multicolumn{1}{c}{---} & $99\pm5$ \\
V1228\,Tau &  $8160\pm300$ & $1.147\pm0.025$ & \multicolumn{1}{c}{---} & $200\pm15$ \\
V650\,Tau & $8300\pm300$ & $1.133\pm0.027$ & $-0.03\pm0.20$ & $230\pm10$ \\
\bottomrule
\end{tabular}
\end{table}

Some of our $\delta$\,Sct stars have received dedicated asteroseismic studies from the ground (e.g., \citealt{lietal2002,suarezetal2002,lietal2004,foxmachadoetal2002,foxmachadoetal2011}). While some of these studies were based on several nights of photometry, the space-based {\it K2} light curves are naturally of higher duty cycle and better precision. All five stars were modelled by \citet{fox-machadoetal2006}, who calculated stellar masses, inclinations and rotation rates. We discuss these in Sec.\,\ref{ssec:lit-models}. First, we describe the properties of the five stars as ascertained from the literature.

We note that these targets will be observed by TESS over three sectors at two-minute cadence, with the full data expected to be released by early 2022.


\subsection{V647\,Tau, $v\sin i = 10$\,km\,s$^{-1}$}

The stellar properties of V647\,Tau have been analysed by many authors. \citet{gasparetal2016} determined the most recent spectroscopic metallicity, [Fe/H] = $0.19 \pm 0.09$, which is in agreement with other Fe abundances, e.g. [Fe/H] = $0.15\pm0.16$ \citep{gebran&monier2008}.
\citet{gebran&monier2008} measured $v\sin i = 18.9$\,km\,s$^{-1}$ from high-resolution spectra, and $T_{\rm eff}$ = 8055\,K using the {\sc uvbybeta} code \citep{napiwotzkietal1993} based on data from \citet{hauck&mermilliod1998}. Elsewhere, smaller $v\sin i$ values with an upper limit of 12\,km\,s$^{-1}$ have been reported \citep{bernacca&perinotto1970,uesugi&fukuda1970,rodriguezetal2000}. We adopt $v \sin i = 10\pm3$\,km\,s$^{-1}$ based on these observations.
Spectroscopic effective temperatures are relatively consistent for V647\,Tau: 8090 and 8100\,K have been determined by \citet{hui-bon-hoa&alecian1998} and \citet{burkhart&coupry1997}. The latter authors also reported $\log_n({\rm Fe})=7.8$, which equates to [Fe/H] = 0.30 under the \citet{asplundetal2009} solar photospheric abundances.

We note that Str\"omgren photometry, via the colour index $b-y = 0.143\pm0.010$\,mag and $\beta=2.793\pm0.041$\,mag \citep{paunzen2015}, suggests considerably cooler temperatures of $T_{\rm eff} = 7150$ and 7300\,K, respectively \citep{murphy2014}. We adopt $T_{\rm eff} = 8080\pm200$\,K, in accordance with spectroscopic results.

There is no indication that this star is in a binary system. RV variability of only 1.0\,km\,s$^{-1}$ exists across three spectra in APOGEE DR16 \citep{ahumadaetal2020}. The star's Renormalised Unit Weight Error in Gaia DR2 is 0.87. 
Objects with RUWE>2.0 are likely to be binaries \citep{evans2018,rizzutoetal2018}.

\subsection{V624\,Tau, $v\sin i = 42$\,km\,s$^{-1}$}

V624\,Tau is also well-studied, with consistent spectroscopic effective temperatures that range from 7600\,K to 8070\,K \citep{burkhart&coupry1997,hui-bon-hoa&alecian1998,gebran&monier2008,kahramanetal2017b}. We adopt a weighted average effective temperature of 7960\,K with a 200\,K uncertainty. Metallicities have also been determined by many studies, with remarkable consistency around [Fe/H] = 0.15 \citep{hui-bon-hoa&alecian1998,gebran&monier2008,gasparetal2016,kahramanetal2017b}, with the exception of the high value of [Fe/H] = $0.4\pm0.1$ from \citet{burkhart&coupry1997}.


Various $v\sin i$ measurements are available, ranging from 32 to 70\,km\,s$^{-1}$. We adopt the median value, $v \sin i = 42\pm5$\,km\,s$^{-1}$, from \citet{solano&fernley1997}.

RV variability of only 2.1\,km\,s$^{-1}$ across three spectra in APOGEE DR16, and an RUWE of 1.03, suggest this star is single.


\subsection{V534\,Tau, $v\sin i = 99$\,km\,s$^{-1}$}


Comparatively fewer studies of V534\,Tau are available. Its rotation rate is consistently measured as between 91 and 99\,km\,s$^{-1}$ \citep{bernacca&perinotto1970,uesugi&fukuda1970,solano&fernley1997}, from which we adopt the most recent measurement of $99\pm5$\,km\,s$^{-1}$. We also adopt the effective temperature from the same source, $T_{\rm eff} = 7600\pm200$\,K, with an assumed uncertainty.

While the RV variability for V534\,Tau is only 2.1\,km\,s$^{-1}$ across three spectra in APOGEE DR16, the RUWE value is extremely high (12.86), perhaps suggesting a binary observed with a face-on orbit.


\subsection{V1228\,Tau, $v\sin i = 200$\,km\,s$^{-1}$}

V1228\,Tau is a rapid rotator, with measurements of 215 and 187\,km\,s$^{-1}$ \citep{bernacca&perinotto1970,uesugi&fukuda1970}. We adopt an intermediate value: $200\pm15$\,km\,s$^{-1}$.


No reliable effective temperature was available in the literature, but Str\"omgren photometry is available in \citet{paunzen2015}. We ran the {\sc uvbybeta} code from the {\sc idlastro} distribution\footnote{\url{https://idlastro.gsfc.nasa.gov/}} on that photometry and obtained $T_{\rm eff} = 8160$\,K and $M_V = 2.27$\,mag. We adopt a 300\,K uncertainty on $T_{\rm eff}$. This value should be treated cautiously, given the difference in $T_{\rm eff}$ using this approach for the stars with high-resolution spectroscopy.

RV variability of 12\,km\,s$^{-1}$ across three spectra in APOGEE DR16 suggests this star might be a binary, though rapid rotation makes precise velocities difficult to determine. The RUWE value (5.71) is also large.

\subsection{V650\,Tau, $v\sin i = 230$\,km\,s$^{-1}$}


This star is also not well-studied, perhaps because of its rapid rotation ($v\sin i = 219$\,km\,s$^{-1}$, \citealt{royeretal2002b}; $240\pm10$\,km\,s$^{-1}$, \citealt{torres2020}). \citet{bochanskietal2018} found [Fe/H] $= -0.03$ from isochrone fitting, but rapid rotation and gravity darkening probably misplace this star with respect to isochrones computed without rotation (see, e.g., \citealt{suarezetal2002}).


Since no effective temperature was available in the literature, we ran the {\sc uvbybeta} code with inputs from \citet{paunzen2015} and obtained $T_{\rm eff} = 8300$\,K and $M_V = 2.05$\,mag. We use a 300\,K uncertainty on $T_{\rm eff}$.

RV variability of 3.1\,km\,s$^{-1}$ is evident between two spectra in APOGEE DR16, but given the rapid rotation and correspondingly large uncertainties, this is not suggestive of binarity, and the RUWE value is also small (1.01). Although this star has been claimed to be RV variable \citep{liuetal1991}, \citet{torres2020} found no significant RV variability in eight high-SNR spectra spanning 594\,d.

\subsection{Stellar luminosities}
\label{ssec:lum}

To parametrise the stars as well as possible with current data, we calculated luminosities using \texttt{ariadne}, a Python package for fitting Spectral Energy Distributions (SEDs).\footnote{\url{https://github.com/jvines/astroARIADNE}} Photometry for the SEDs came from a combination of the \textit{ALL-WISE} \citep{Lang2016WISE}, \textit{APASS} \citep{2016yCat.2336....0H}, \textit{Pan-STARRS1} \citep{2016arXiv161205560C}, \textit{2MASS} \citep{Skrutskie2006Two}, and \textit{Tycho-2} \citep{Hog2000Tycho2} catalogues where data was available. Hence, the wavelength range spans from about 0.4 to 25\,$\upmu$m, but not all stars had data in all bands.
Temperatures were constrained following the values in Table~\ref{tab:identifiers}, and parallaxes were preferentially chosen from either the \textit{Gaia} eDR3 \citep{Collaboration2021Gaia} or DR2 \citep{GaiaCollaboration2018Gaia} catalogues, depending on the associated uncertainty. The models were fitted using {\small ATLAS9} \citep{castelli&kurucz2003}, which was chosen because it most accurately reproduces the bluest colours \citep{martins&coelho2007}. Each SED model was sampled with the \textsc{PyMultiNest} nested sampling package until convergence, defined as a change in the log probability less than 0.1 \citep{Buchner2014Xray}. We summarise the obtained luminosities in Table~\ref{tab:obs}.

\subsection{Models in the literature}
\label{ssec:lit-models}

Stellar parameters are also available from asteroseismology from the study by \citet{fox-machadoetal2006}. They modelled six $\delta$\,Sct stars from the Pleiades, including all five of our targets plus HD\,23194 (V1187\,Tau), which was not observed with K2. Their models included rotation and are the most homogeneous set of asteroseismic results in the literature. We provide their inferred stellar parameters in Table\:\ref{tab:lit_models}. While their methodology was well-conceived, relatively few modes were clearly detectable from ground-based photometry, leading to fewer constraints on the models. Many pulsation modes were therefore ambiguously identified, with a total of 47 identifications for 29 observed modes. The imposition of a hard age prior that was too young (70--100\,Myr; cf. 100--160\,Myr, Sec.\,\ref{sec:intro}) might also have affected the results.

\begin{table}
\centering
\caption{Stellar parameters from the modelling of \citet{fox-machadoetal2006}, based on a best-fitting cluster metallicity of [Fe/H] = 0.067 using $(Z/X)_{\sun} = 0.0245$ from \citet{grevesse&noels1993}, core overshooting in the range $0.0 < \alpha_{\rm ov} < 0.2$, an age of 70-100\,Myr, and a distance modulus of 5.6--5.7.}
\label{tab:lit_models}
\begin{tabular}{lccc}
\toprule
Name 	&	\multicolumn{1}{c}{M}	&	\multicolumn{1}{c}{$\nu_{\rm rot}$}	&	\multicolumn{1}{c}{$i$}	\\
(GCVS)	&	M$_{\sun}$	&	\multicolumn{1}{c}{$\upmu$\,Hz}	&	\multicolumn{1}{c}{deg}	\\
\midrule
V647\,Tau &	1.68--1.72	&	10--11	&	17--18	\\
V624\,Tau &	1.68--1.72	&	3--5	&	37--67	\\
V534\,Tau &	1.65--1.69	&	14--16	&	59--79	\\
V1228\,Tau &	1.82--1.86	&	24--26	&	53--59	\\
V650\,Tau &	1.84--1.88	&	25--28	&	60--70	\\
\bottomrule
\end{tabular}
\end{table}

\section{K2 light curves}
\label{sec:k2}

\begin{figure*}
\begin{center}
\includegraphics[width=0.7\textwidth]{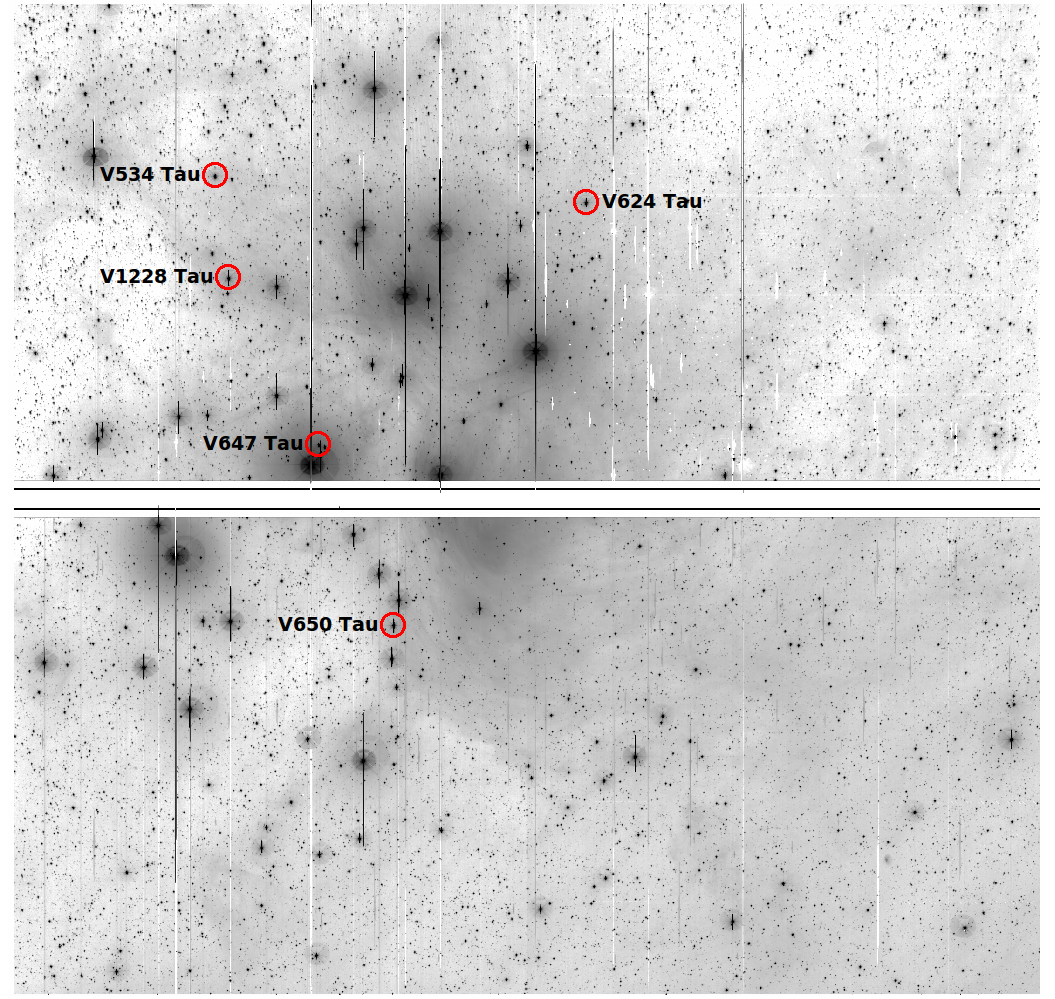}
\includegraphics[width=0.24\textwidth]{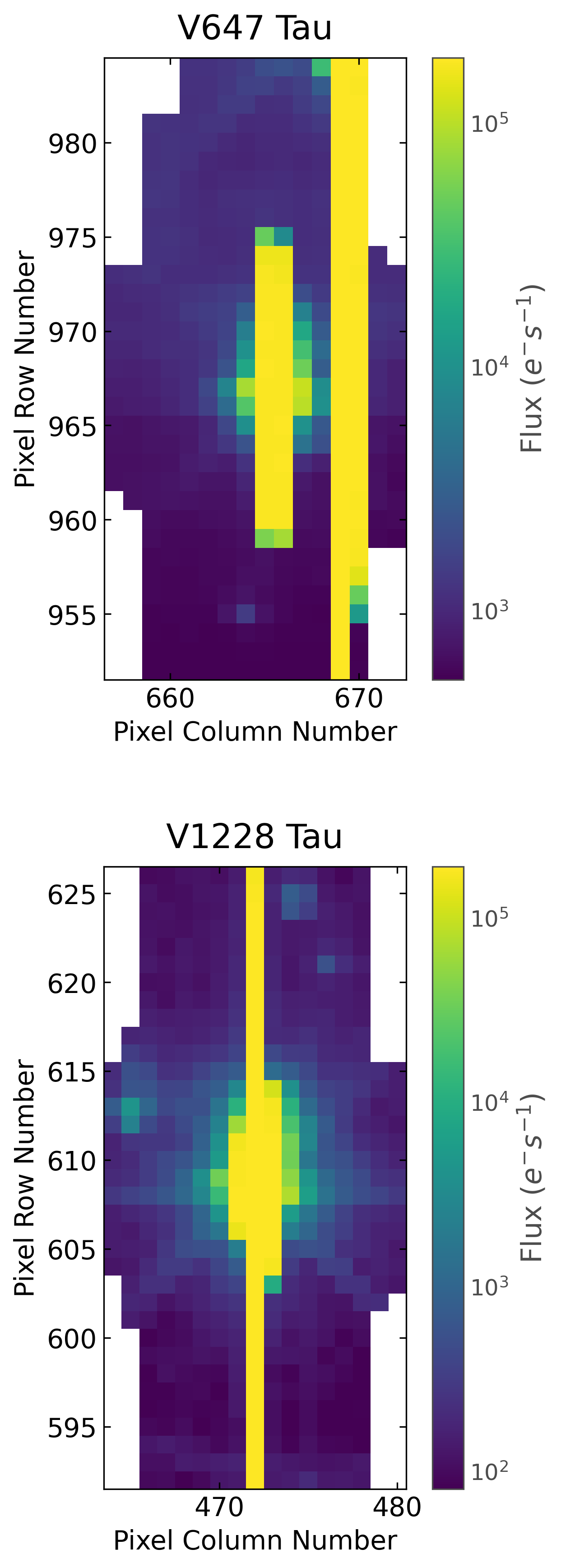}
\caption{\emph{Left:} Full frame image of Kepler CCD Module 15 from K2 Campaign 4. The locations of the five $\delta$\,Sct stars are circled. \emph{Top right:} Single frame image of V647\,Tau from its target pixel file. The bleed column from the nearby star 24\,Tau also passes through the image. \emph{Bottom right:} Single frame image of V1228\,Tau from its target pixel file.}
\label{fig:ffi}
\end{center}
\end{figure*}

The Pleiades were observed during K2 Campaign 4 between 2015 February 8 and 2015 April 20. 
The brightest seven stars, of spectral type B, were analysed by \citet{whiteetal2017}.
In this paper we study the five previously-known $\delta$\,Sct stars that were targeted in short-cadence mode. Fig.\,\ref{fig:ffi} shows the full-frame image of Kepler CCD Module 15, upon which the majority of the Pleiades cluster fell. The locations of our targets are indicated.

During the K2 Mission, the Kepler spacecraft had only two functioning reaction wheels, when three are necessary to maintain stable pointing along three axes. To maintain relatively stable pointing, the spacecraft was aligned to observe fields in the ecliptic because this minimized the torque due to solar radiation pressure about the roll axis. Gradual drifts in the pointing were adjusted by thruster firings at approximately 6\,h intervals \citep{howelletal2014}.

As a consequence of this observing mode, K2 light curves generated from the Kepler Mission Data Processing Pipeline \citep{jenkinsetal2010b,jenkins2020} typically contain strong pointing-drift and thruster-firing systematics. Several pipelines have been developed to further process K2 data and remove these systematic errors, with corrected light curves available as high-level science products through the Mikulski Archive for Space Telescopes \citep{vanderburg&johnson2014,aigrainetal2015,aigrainetal2016,armstrongetal2016,barrosetal2016,libralato2016,lugeretal2016,lugeretal2018,whiteetal2017,popeetal2019}. However, with few exceptions, these data products are only available for long-cadence data and, consequently, the only short-cadence light curves presently available for our targets through the MAST are from the original Kepler pipeline.

\begin{figure*}
\begin{center}
\includegraphics[width=0.98\textwidth]{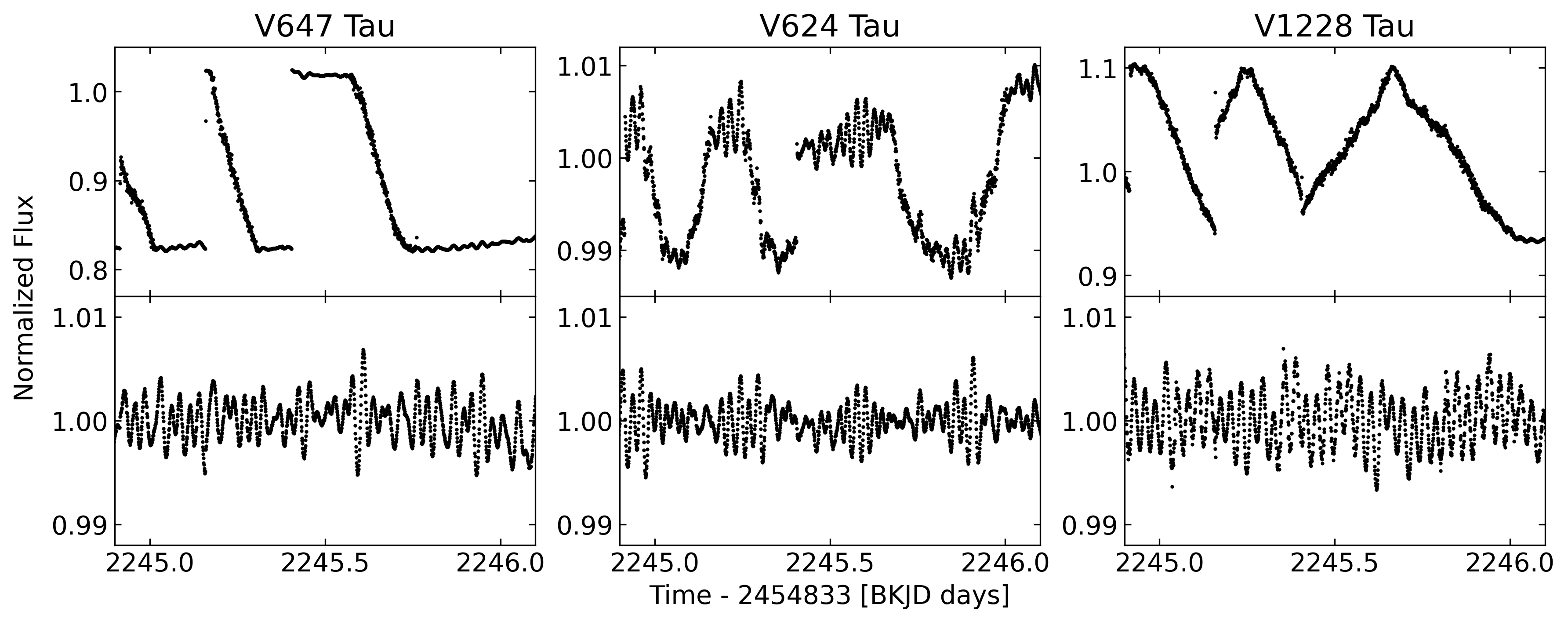}
\caption{Segments of the light curves for three of the $\delta$\,Sct stars. The top row shows the light curves provided by the Kepler pipeline. The bottom row shows our corrected light curves. Note the change in scale of the y-axis.}
\label{fig:lcs}
\end{center}
\end{figure*}

The Kepler pipeline light curves for our five stars vary greatly in quality. For V534\,Tau and V650\,Tau, the pointing systematics are negligible relative to the $\delta$ Sct pulsations and we used these pipeline light curves in our analysis. 
For the other three stars, however, the pointing systematics in the light curves are substantial. Segments of these light curves are shown in the top row of Fig.~\ref{fig:lcs}. For these stars we have generated our own light curves from the pixel-level data using the \textsc{lightkurve} Python package \citep{lightkurvecollaboration2018}. Each of these stars presented unique challenges and required an individual treatment, which we describe in the following sections. The light curves, and \textsc{jupyter} notebooks used to generate them, are available at \url{https://github.com/hvidy/k2-pleaides-dsct}.

\subsection{Light curve for V647 Tau}
V647\,Tau is located in a relatively crowded region of the cluster, which complicates light curve creation for this star. The bleed column from the saturated star 24\,Tau passes within $\sim$4 pixels of V647\,Tau, as can be seen in Fig.~\ref{fig:ffi}. Due to the pointing drift, any chosen aperture will either experience incursions of this bleed column, resulting in jumps in flux, or have significant aperture losses when V647\,Tau partially drifts out of the aperture.

To overcome this problem, we generated two light curves, one occasionally affected by the bleed column of 24\,Tau and one occasionally affected by aperture losses. Importantly, these problems do not occur simultaneously. We were therefore able to generate a merged light curve from sections of both light curves that are not affected by either problem. A segment of this light curve is shown in the bottom left panel of Fig.~\ref{fig:lcs}.

\subsection{Light curve for V624 Tau}
The pointing systematics present in the Kepler pipeline light curve of V624\,Tau are mostly a result of an insufficiently large aperture being used. As a result, there are losses when some of the flux from the target drifts outside the aperture.
We determined new aperture masks that were large enough to avoid significant aperture losses. Because the pointing drift behaviour varies over the course of the observations, we split the light curve into three segments and chose an appropriate mask for each.
The light curves generated with our new aperture masks still exhibit some pointing systematics. We removed these using the `Self Flat Fielding' (SFF) method of \citet{vanderburg&johnson2014}, as implemented in the \textsc{lightkurve} package. A segment of the corrected light curve is shown in the bottom middle panel of Fig.~\ref{fig:lcs}.

\subsection{Light curve for V1228 Tau}
As for V624\,Tau, the pointing systematics present in the Kepler pipeline light curve are largely a result of the aperture mask not being large enough. Unlike for V624\,Tau, however, this situation could not be simply rectified by choosing more appropriate aperture masks. This is because the target pixel file itself is not large enough. Depending on the position of the star, some flux bleeds off the edge of the target pixel file and is lost, as can be seen in Fig.~\ref{fig:ffi}. 

To create the light curve for V1228\,Tau, we first selected new apertures that minimise the aperture losses as much as possible. The remaining aperture losses do correlate with the position of the star on the CCD, but in a complex way such that the SFF method in \textsc{lightkurve} was not able to adequately detrend the light curve. We circumvented this by identifying which data segments showed similar, simple correlations between the star's CCD position and aperture loss, and detrending these segments together. A section of the final corrected light curve is shown in the bottom right panel of Fig.~\ref{fig:lcs}.

\section{Pulsation spectra and mode identification}
\label{sec:pulsations}

\begin{figure*}
\begin{center}
\includegraphics[width=0.98\textwidth]{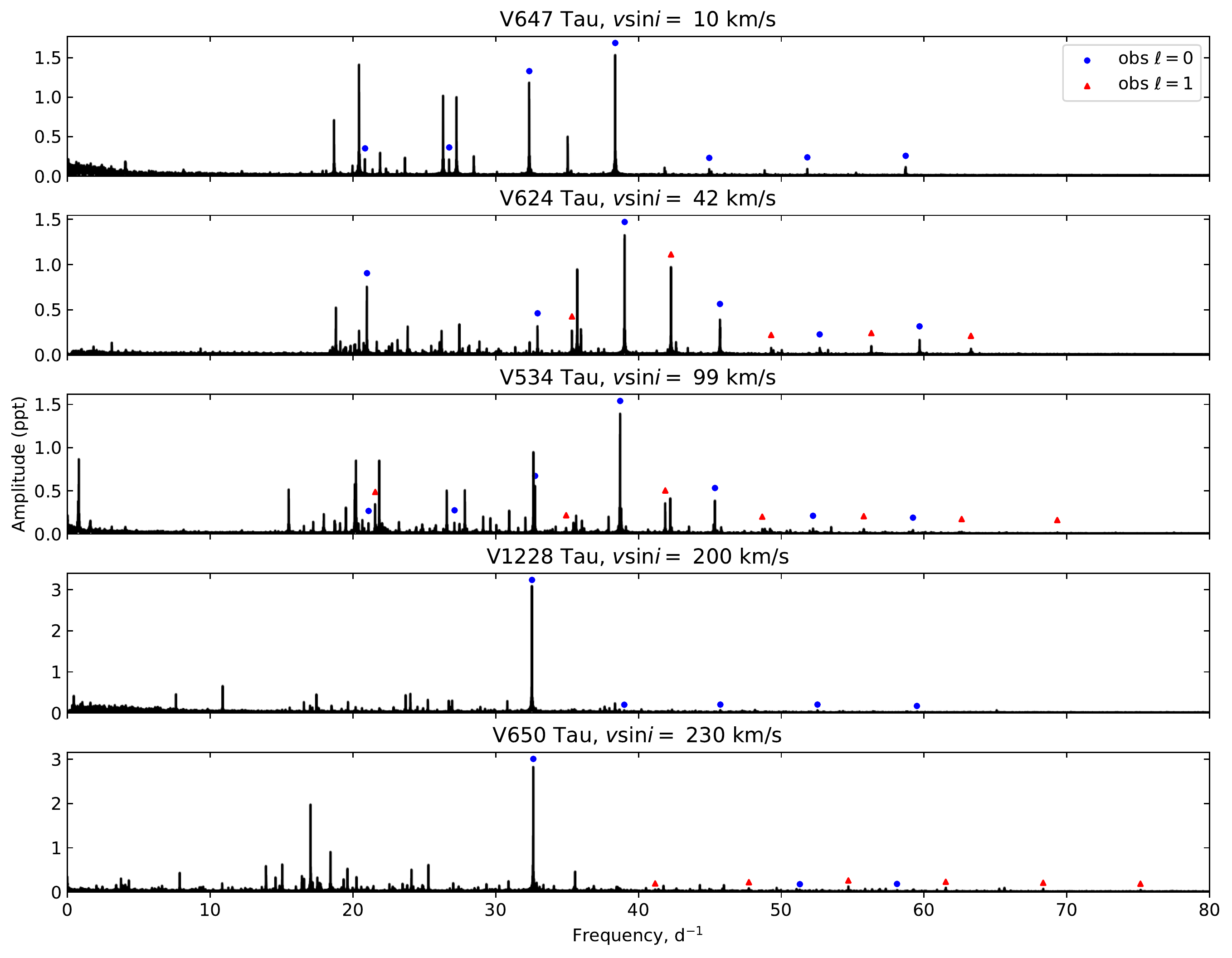}
\caption{Fourier amplitude spectra of the five $\delta$\,Sct stars, ordered by $v\sin i$. These are shown in \'echelle format in Fig.\,\ref{fig:ech}. Blue circles indicate peaks identified as $\ell=0$ modes and red triangles are $\ell=1$ modes (see Sec.\,\ref{sec:pulsations}).}
\label{fig:fts}
\end{center}
\end{figure*}

\begin{figure*}
\begin{center}
\includegraphics[width=0.33\textwidth]{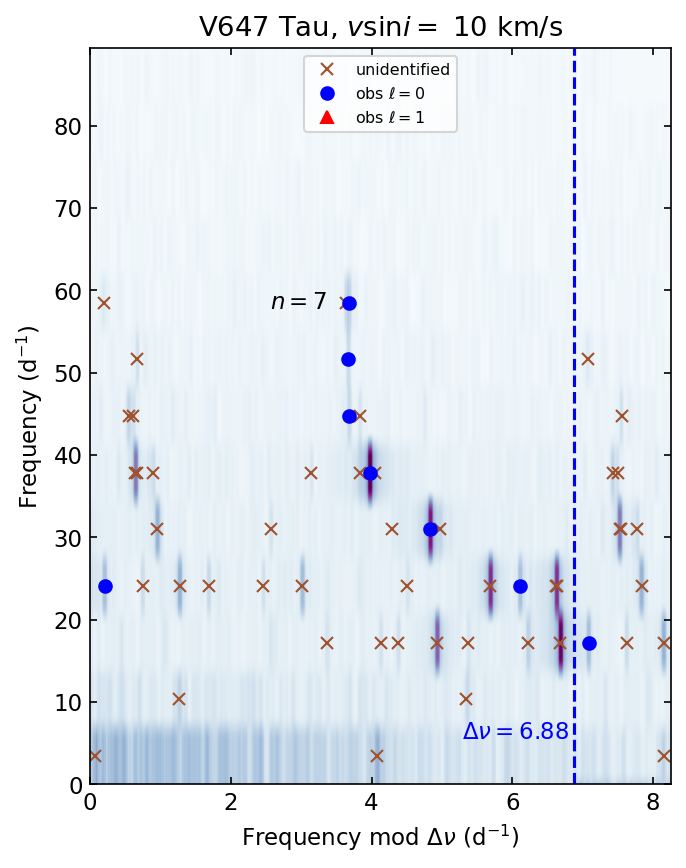}
\includegraphics[width=0.33\textwidth]{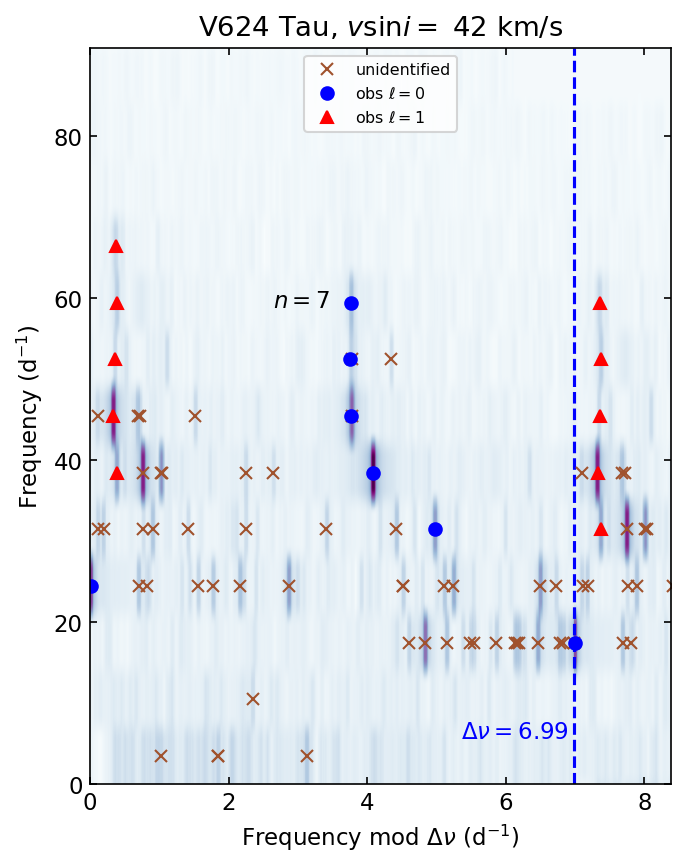}
\includegraphics[width=0.33\textwidth]{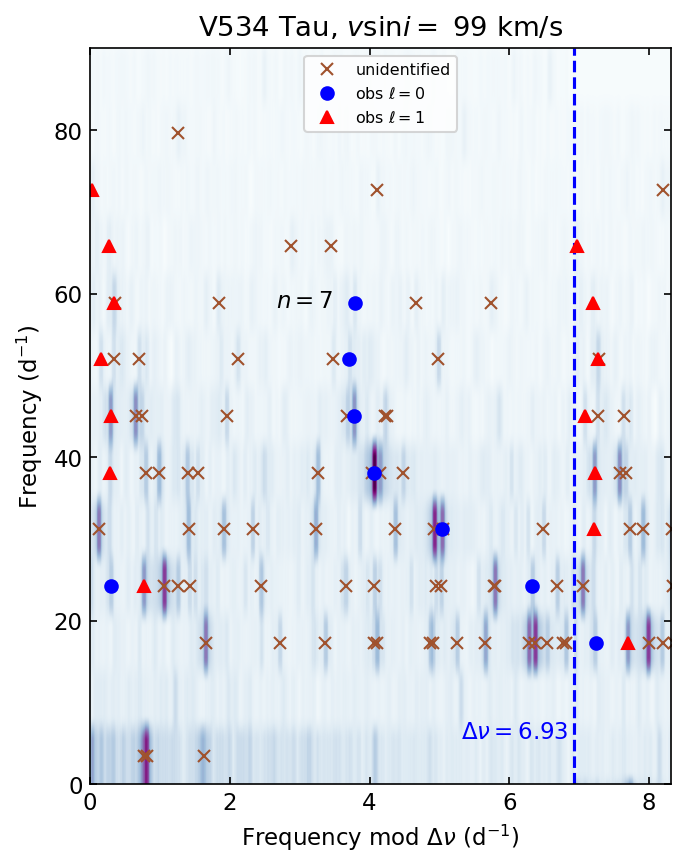}\\
\includegraphics[width=0.33\textwidth]{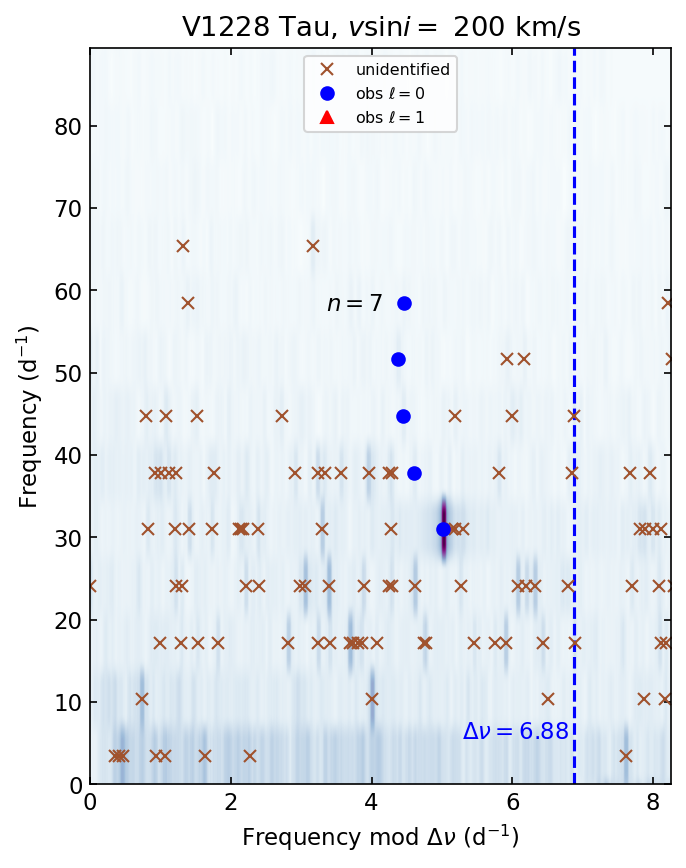}
\includegraphics[width=0.33\textwidth]{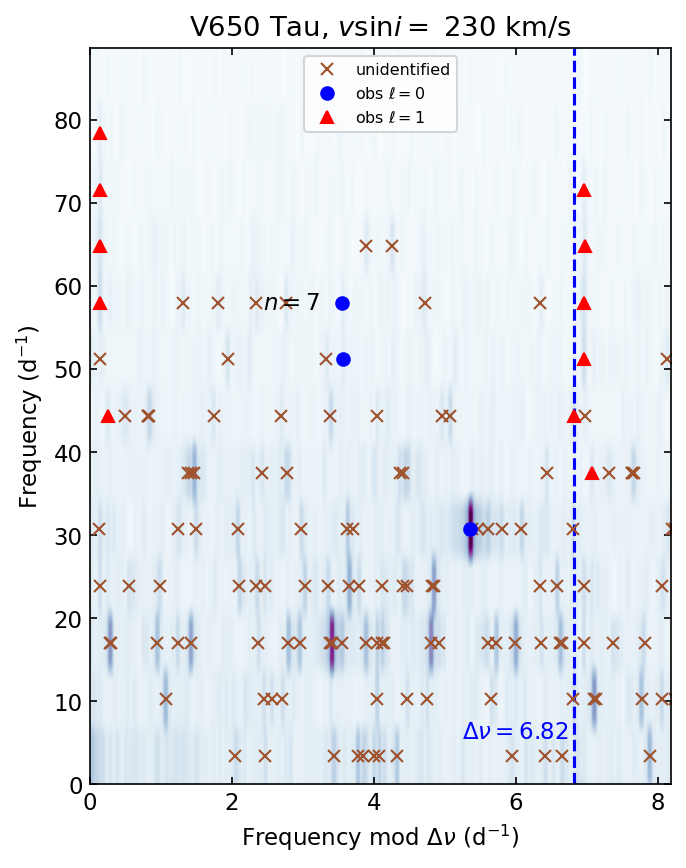}
\caption{\'Echelle diagrams for the five $\delta$\,Sct stars showing the mode identifications.}
\label{fig:ech}
\end{center}
\end{figure*}

The Fourier amplitude spectra for all five stars are shown in Fig.\,\ref{fig:fts}.  In Fig.\,\ref{fig:ech} we show them in \'echelle format, where the amplitude spectra have been divided into segments of equal length and stacked vertically.  As found by \citet{beddingetal2020}, in this class of high-frequency $\delta$\,Sct stars, some of the modes tend to fall along two ridges in the \'echelle diagram, which correspond to $\ell=0$ and $\ell=1$.  These identified peaks are marked by blue circles and red triangles, respectively.  In the following sections we discuss the mode identifications for each of the stars in more detail. We only analyse p\:modes in this work. Searching for and identifying high-order g\:modes like those of $\gamma$\,Dor variables generally requires longer time-series than one K2 campaign.

\subsection{Mode identification for V647\,Tau}
\label{ssec:v647}

\begin{figure*}
\begin{center}
\includegraphics[width=0.33\textwidth]{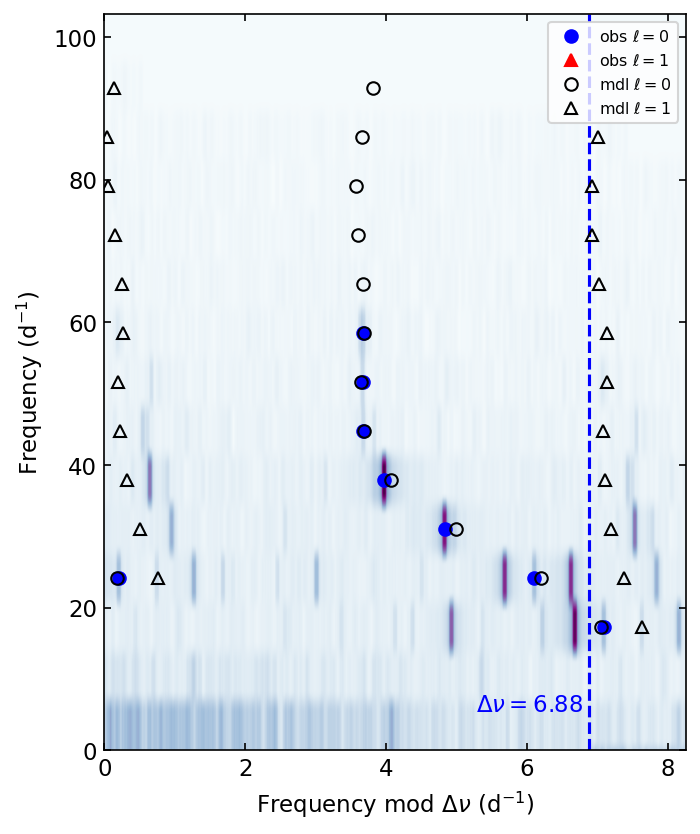}
\includegraphics[width=0.33\textwidth]{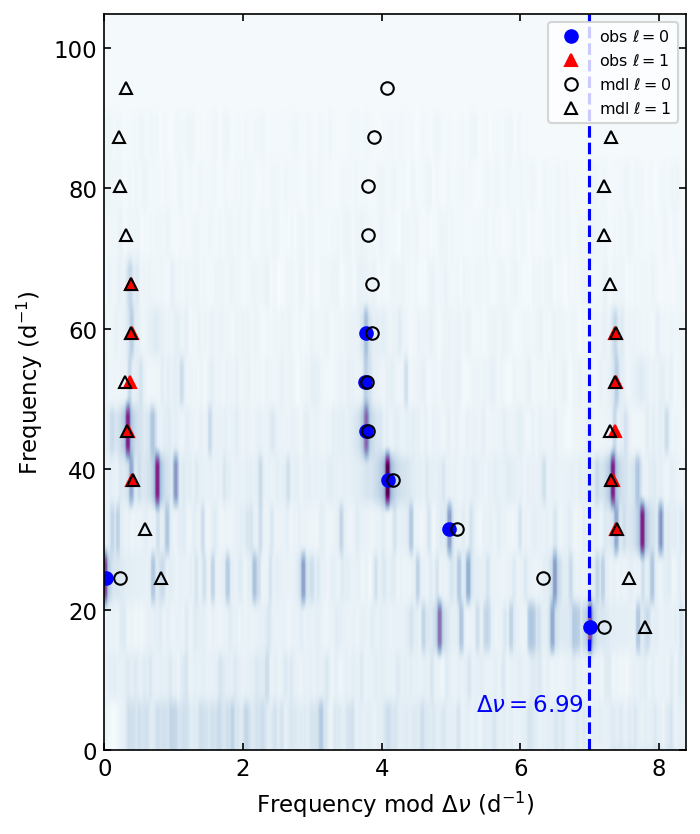}
\includegraphics[width=0.33\textwidth]{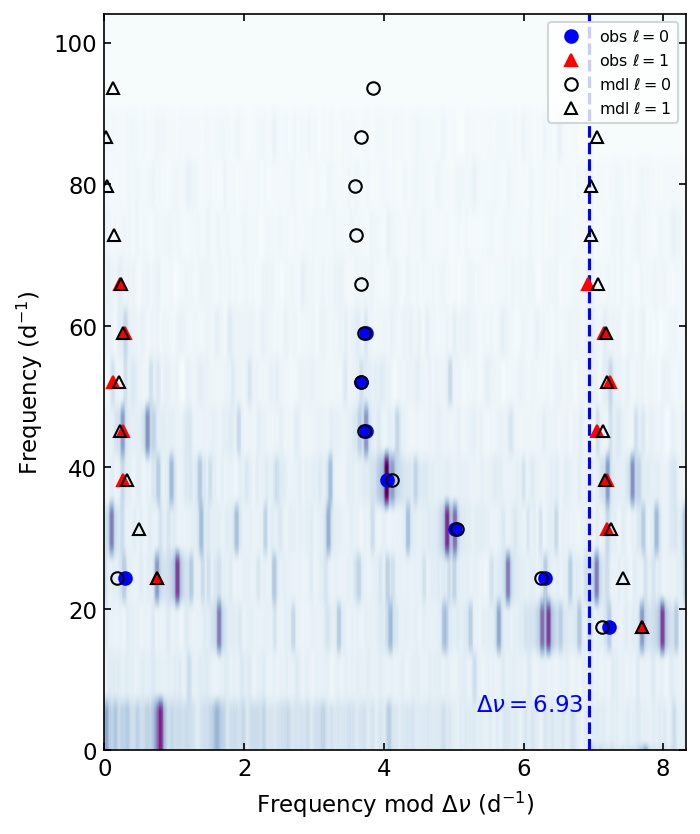}
\caption{\'Echelle diagrams with models for the three $\delta$\,Sct stars with the most secure mode identifications. {\bf Left:} V647 Tau, {\bf center:} V624 Tau, {\bf right:} V534 Tau. For reference, the lowest frequency $\ell=0$ mode in each case has $n=1$, and the highest frequency $\ell=0$ mode identified in each case has $n=7$. The models are described in Sec.\,\ref{sec:models}.}
\label{fig:ech_model}
\end{center}
\end{figure*}

The \'echelle diagram of V647\,Tau shows a strong radial ridge curving through the centre (Fig.\,\ref{fig:ech}; filled blue circles). While the higher-order radial modes are unambiguous, the bottom two rows ($n=2,1$) have some strong modes that we have not identified as the radial modes. Our identifications were aided by the calculation of frequency ratios \citep{suarezetal2006,netzeletal2021,yangetal2021} and by comparison with non-rotating models (Fig.\,\ref{fig:ech_model}), whose calculation we describe in Sec.\,\ref{sec:models}.  The observed radial mode frequencies are given in Table\:\ref{tab:IDs}.

For the case of V647\,Tau, we have not identified modes in the dipole ridge (in other stars these are shown as filled red triangles) because the peaks on the left side of the \'echelle fall along a tilted ridge that sits offset to higher frequency than the non-rotating model frequencies in Fig.\,\ref{fig:ech_model}. This suggests that this observed ridge comprises prograde ($m=+1$) dipole modes, with the zonal ($m=0$) modes not being detected. We do not know the inclination of the rotation axis, or understand mode selection in $\delta$\,Sct stars well enough in general, to explain why only a prograde ridge is observed. It may be relevant that for the high-order g\:modes observed in $\gamma$\,Dor stars, more rapid rotation tends to favour the excitement of prograde modes \citep{glietal2020a}, but if V647\,Tau were a rapid rotator then its low $v \sin i$ implies it is seen seen pole-on and the visibility of prograde modes of any kind should then be low \citep{aertsetal2010,reeseetal2013}.

At low rotation rates, rotational splittings should be approximately symmetrical. Peaks that lie equidistant from the non-rotating models of $\ell=1$ $m=0$ modes could therefore comprise a rotationally split pair. On this basis, and by assuming that the observed $\ell=1$ ridge comprises prograde modes, we can speculate on possible additional mode identifications. The strong peak lying just to the right of the $n=2$ radial mode, at 27.263\,d$^{-1}$, could be the $\ell=1$ retrograde mode on that order, or the strong unidentified peak at 20.439\,d$^{-1}$, lying just to the left of the fundamental radial mode, might be a retrograde $\ell=1$ mode. The long $\ell=1$ prograde ridge, in particular, should allow an asteroseismic rotation rate to be derived in future work using rotating models. 

\subsection{Mode identification for V624\,Tau}
\label{ssec:v624}

The \'echelle diagram of V624\,Tau shows a strong radial ridge curving through the centre (Fig.\,\ref{fig:ech}), with less ambiguity at lower radial orders than with V647\,Tau, although we have not identified the $n=2$ radial mode. While the strong peak at 27.464\,d$^{-1}$ has almost the correct frequency, its ratios with the $n=1$ and $n=3$ modes are significantly different from the expected values. Specifically, the difference between the observed and model frequency ratios, $f_{n=1}/f_{n=3}$ is only 0.0040, whereas the ratio for the $n=1$ and (unlabelled) $n=2$ modes differs by 0.0121.

The $\ell=1$ ridge in V624\,Tau appears to comprise $m=0$ modes, matching well with non-rotating models (Fig.\,\ref{fig:ech_model}). The $n=2$ peak is absent, and the $n=1$ peak is ambiguous between two relatively weak peaks near 21.7\,d$^{-1}$. We note that even the slight weave in the observed dipole ridge is well-reproduced by the models.

\subsection{Mode identification for V534\,Tau}
\label{ssec:v534}

The $\ell=1$ zonal modes are readily identifiable in V534\,Tau, but precise identification of the radial modes is difficult because at $n=5, 4$ and 3 there are two very closely spaced peaks. In each of these rows, one peak will be the radial mode while the other is probably a prograde $\ell=2$ mode. At $n=3$ the mode amplitudes are similar, so we identified the lower-frequency peak as the radial mode, since this gives a better fit to the expected frequency ratios from non-rotating models. At $n=4$ and 5, the amplitudes of the close peaks are somewhat different and we chose the stronger peak of each pair to be the radial mode. Visibility curves of rotating stars suggest that radial modes should be $\sim$2--5 times stronger than prograde $\ell=2$ modes at $n=4$ \citep{reeseetal2013}, based on loosely-informed assumptions of an inclination of 75$^{\circ}$ \citep{fox-machadoetal2006} and a rotation rate of 20\% of the critical value. These doublets and the many peaks between the $\ell=0$ and 1 ridges suggest that many rotationally-split $\ell=2$ multiplets ought to be found in rotating models of this star.

We note that, as with V647\,Tau, the $n=1$ and $n=2$ radial modes appear to be the weaker of the possible peaks on their rows. The frequencies modulo $\Delta\nu$ of these modes are similar in both stars. Our identifications of all $m=0$ modes are given in Table\:\ref{tab:IDs}.

With the $m=0$ modes accounted for, the majority of the remaining strong peaks must be rotationally split modes. The visual similarity of the amplitude spectra and the \'echelles of V647\,Tau and V534\,Tau, especially around the dipole ridge, therefore suggests that these stars have similar interior rotation rates, and that their differences in measured $v \sin i$ are a product of different observed inclination angles.

\subsection{Mode identification for V1228\,Tau}
\label{ssec:v1228}

Mode identification in V1228\,Tau and V650\,Tau is less straightforward due to their more rapid rotation. Looking at the oscillation spectra of all five stars as an ensemble (Fig.\,\ref{fig:fts}), we suggest that the strongest peak in the two rapid rotators corresponds to the $n=3$ radial mode.
This identification helps to choose the correct $\Delta\nu$ for the \'echelle diagrams. This approach was not necessary for the slower rotators, whose \'echelle diagrams only show vertical ridges at specific values of $\Delta\nu$ that also place their strongest modes firmly on the radial ridge.

Three weaker radial modes at $n=5$--7 are evident for V1228\,Tau, and their identification aids in selecting the correct $n=4$ mode. At lower orders, identification of the radial modes is complicated by three close peaks of similar amplitude at $n=2$, and only weak peaks at $n=1$. Dipole modes could not be readily identified. We do not plot model frequencies for this star because non-rotating models cannot be expected to be accurate for a star with $v \sin i=200$\,km\,s$^{-1}$. Our mode identifications should be considered particularly tentative for this star.

\subsection{Mode identification for V650\,Tau}
\label{ssec:v650}

Two key features anchor the mode identifications for this star. Firstly, the strong $n=3$ radial mode at a similar frequency to those of the other four stars, and secondly, a weak but relatively straight dipole ridge of six peaks, stretching to $n=9$. Based on these anchor points, frequencies of 51.303 and 58.108\,d$^{-1}$ are tentatively suggested for the $n=6$ and 7 radial modes. We note that the identification of the $n=5$ dipole mode is uncertain, because selecting the weaker and higher-frequency option of a close pair would result in a straighter ridge. Nonetheless, we elected for the stronger peak. As with V1228\,Tau, due to the inadequacies of our non-rotating models, these identifications should be considered tentative. Finally, we note that the number of excited modes appears to be much higher in the two rapid rotators than in the two slowest rotators.

\begin{table*}
\centering
\caption{Identified modes for the five Pleiades $\delta$\,Sct stars. Results for the two rapid rotators, V1228\,Tau and V650\,Tau, should be considered tentative. All modes have $m=0$.}
\label{tab:IDs}
\begin{tabular}{cccccccccccccc}
\toprule
\multicolumn{4}{c}{\bf V647\,Tau} && \multicolumn{4}{c}{\bf V624\,Tau} && \multicolumn{4}{c}{\bf V534\,Tau} \\
\midrule            
	\multicolumn{1}{c}{$f_{\rm obs}$}	&	Amp.	&	$n$	&	$\ell$	&&	\multicolumn{1}{c}{$f_{\rm obs}$}	&	Amp.	&	$n$	&	$\ell$	&&	\multicolumn{1}{c}{$f_{\rm obs}$}	&	Amp.	&	$n$	&	$\ell$	\\
	\multicolumn{1}{c}{d$^{-1}$}	&	mmag	& & &&	\multicolumn{1}{c}{d$^{-1}$}	&	mmag	& & &&	\multicolumn{1}{c}{d$^{-1}$}	&	mmag	& & \\\midrule
$ 20.8493 $&$ 0.202 $&$ 1 $&$ 0 $&$ \quad\quad$&$ 20.9829 $&$ 0.755 $&$ 1 $&$ 0 $&$ \quad\quad$&$ 21.0945 $&$ 0.118 $&$ 1 $&$ 0 $\\
$ 26.7418 $&$ 0.214 $&$ 2 $&$ 0 $&$ \quad\quad$&$ 32.9335 $&$ 0.312 $&$ 3 $&$ 0 $&$ \quad\quad$&$ 27.1158 $&$ 0.126 $&$ 2 $&$ 0 $\\
$ 32.3502 $&$ 1.180 $&$ 3 $&$ 0 $&$ \quad\quad$&$ 39.0318 $&$ 1.320 $&$ 4 $&$ 0 $&$ \quad\quad$&$ 32.7572 $&$ 0.523 $&$ 3 $&$ 0 $\\
$ 38.3728 $&$ 1.536 $&$ 4 $&$ 0 $&$ \quad\quad$&$ 45.7058 $&$ 0.415 $&$ 5 $&$ 0 $&$ \quad\quad$&$ 38.7159 $&$ 1.390 $&$ 4 $&$ 0 $\\
$ 44.9568 $&$ 0.081 $&$ 5 $&$ 0 $&$ \quad\quad$&$ 52.6896 $&$ 0.079 $&$ 6 $&$ 0 $&$ \quad\quad$&$ 45.3571 $&$ 0.383 $&$ 5 $&$ 0 $\\
$ 51.8260 $&$ 0.088 $&$ 6 $&$ 0 $&$ \quad\quad$&$ 59.6845 $&$ 0.168 $&$ 7 $&$ 0 $&$ \quad\quad$&$ 52.2233 $&$ 0.062 $&$ 6 $&$ 0 $\\
$ 58.7160 $&$ 0.107 $&$ 7 $&$ 0 $&$ \quad\quad$&$ 35.3390 $&$ 0.278 $&$ 3 $&$ 1 $&$ \quad\quad$&$ 59.2264 $&$ 0.040 $&$ 7 $&$ 0 $\\
$ \phantom{21.9161} $&$ \phantom{0.288} $&$ \phantom{1} $&$ \phantom{1} $
&$ \quad\quad$&$ 42.2781 $&$ 0.963 $&$ 4 $&$ 1 $&$ \quad\quad$&$ 21.5616 $&$ 0.338 $&$ 1 $&$ 1 $\\
$ \phantom{28.4751} $&$ \phantom{0.244} $&$ \phantom{2} $&$ \phantom{1} $
&$ \quad\quad$&$ 49.2938 $&$ 0.074 $&$ 5 $&$ 1 $&$ \quad\quad$&$ 34.9341 $&$ 0.069 $&$ 3 $&$ 1 $\\
$ \phantom{35.0464} $&$ \phantom{0.501} $&$ \phantom{3} $&$ \phantom{1} $
&$ \quad\quad$&$ 56.3099 $&$ 0.094 $&$ 6 $&$ 1 $&$ \quad\quad$&$ 41.8765 $&$ 0.355 $&$ 4 $&$ 1 $\\
$ \phantom{41.8306} $&$ \phantom{0.106} $&$ \phantom{4} $&$ \phantom{1} $
&$ \quad\quad$&$ 63.2843 $&$ 0.064 $&$ 7 $&$ 1 $&$ \quad\quad$&$ 48.6669 $&$ 0.052 $&$ 5 $&$ 1 $\\
$ \phantom{48.8315} $&$ \phantom{0.070} $&$ \phantom{5} $&$ \phantom{1} $
&$ \quad\quad$&$  $&$  $&$  $&$  $&$ \quad\quad$&$ 55.7788 $&$ 0.058 $&$ 6 $&$ 1 $\\
$  $&$  $&$  $&$  $&$ \quad\quad$&$  $&$  $&$  $&$  $&$ \quad\quad$&$ 62.6362 $&$ 0.025 $&$ 7 $&$ 1 $\\
$  $&$  $&$  $&$  $&$ \quad\quad$&$  $&$  $&$  $&$  $&$ \quad\quad$&$ 69.3341 $&$ 0.013 $&$ 8 $&$ 1 $\\
\midrule
     \multicolumn{4}{c}{\bf V1228\,Tau}    &&     \multicolumn{4}{c}{\bf V650\,Tau}   &&   &  &  &  \\
\midrule
 $f_{\rm obs}$ & Amp. & $n$ & $\ell$ & & $f_{\rm obs}$ & Amp. & $n$ & $\ell$ &&   &  &  &  \\
 d$^{-1}$ & mmag &  &  & & d$^{-1}$ & mmag &  &  &&   &  &  &  \\
\midrule                          
$ 32.5410 $&$ 3.086 $&$ 3 $&$ 0 $&$ \quad\quad$&$ 32.6333 $&$ 2.859 $&$ 3 $&$ 0 $&$   $&$  $&$  $&$  $\\
$ 39.0053 $&$ 0.053 $&$ 4 $&$ 0 $&$ \quad\quad$&$ 51.3028 $&$ 0.028 $&$ 6 $&$ 0 $&$   $&$  $&$  $&$  $\\
$ 45.7334 $&$ 0.057 $&$ 5 $&$ 0 $&$ \quad\quad$&$ 58.1077 $&$ 0.034 $&$ 7 $&$ 0 $&$   $&$  $&$  $&$  $\\
$ 52.5370 $&$ 0.058 $&$ 6 $&$ 0 $&$ \quad\quad$&$ 41.1683 $&$ 0.046 $&$ 4 $&$ 1 $&$   $&$  $&$  $&$  $\\
$ 59.5018 $&$ 0.023 $&$ 7 $&$ 0 $&$ \quad\quad$&$ 47.7311 $&$ 0.075 $&$ 5 $&$ 1 $&$   $&$  $&$  $&$  $\\
$  $&$  $&$  $&$  $&$ \quad\quad$&$ 54.6997 $&$ 0.114 $&$ 6 $&$ 1 $&$   $&$  $&$  $&$  $\\
$  $&$  $&$  $&$  $&$ \quad\quad$&$ 61.5232 $&$ 0.084 $&$ 7 $&$ 1 $&$   $&$  $&$  $&$  $\\
$  $&$  $&$  $&$  $&$ \quad\quad$&$ 68.3472 $&$ 0.062 $&$ 8 $&$ 1 $&$   $&$  $&$  $&$  $\\
$  $&$  $&$  $&$  $&$ \quad\quad$&$ 75.1600 $&$ 0.040 $&$ 9 $&$ 1 $&$   $&$  $&$  $&$  $\\
\bottomrule
\end{tabular}
\end{table*}

\section{Stellar Models}
\label{sec:models}

\subsection{Model Calculation}

To evaluate our mode identifications, we calculated some provisional, non-rotating models of the five $\delta$\,Sct stars using the latest stable release versions of {\sc mesa} (r15140, \citealt{paxtonetal2011,paxtonetal2013,paxtonetal2015,paxtonetal2018,paxtonetal2019}) and {\sc gyre} (v6.0.1, \citealt{townsend&teitler2013}). 

Evolutionary tracks were calculated in a two-dimensional grid of mass and metallicity. Masses ranged from 1.6 to 1.9\,M$_{\odot}$ in steps of 0.01\,M$_{\odot}$. We used three values for the initial metal mass fraction, $Z_{\rm in} =$ 0.0136, 0.0152, and 0.0171. These correspond to [Fe/H] $= +0.03\pm0.05$, which is the Pleiades metallicity from \citet{soderblometal2009} converted using the \citet{asplundetal2009} bulk solar abundances, which have $Z_{\odot} = 0.0142$. We calculated helium mass fractions, $Y_{\rm in}$, using the following formulae:\\
\begin{equation}
\frac{{\rm d}Y}{{\rm d}Z} = \frac{Y_{\rm in} - Y_0}{Z_{\rm in} - Z_{\odot}} = 1.4, {\rm and}
\label{eq:dydz}
\end{equation}
\begin{equation}
X + Y + Z = 1,
\end{equation}
with $Y_0 = 0.280$. Thus, rather than using the solar helium abundance, we allowed for some evolution of the Galactic helium abundance in the 4.57\,Gyr since the birth of the Sun \citep{balser2006, casagrandeetal2007,chaussidon2007,bouvier&wadhwa2010,silvaaguirreetal2017}. 
We set the initial abundance ratios of $^2$H/$^1$H and $^3$He/$^4$He as $2.0 \times 10^{-5}$ and $1.66 \times 10^{-4}$, respectively \citep{stahleretal1980,linsky1998}. Nuclear burning was calculated using the {\tt pp\_and\_cno\_extras} nuclear net. No elemental diffusion was included in our calculations.

Our models included the pre-main sequence, starting with an initial central temperature of $9\times10^5$\,K. This is higher than the default value of $3\times10^5$\,K, to help with model convergence, but this change makes no substantial difference to the inferred stellar ages (\citealt{soderblom2010}; see also the discussion in \citealt{murphyetal2021a}).

We used fixed-opacity, {\tt Eddington} {\tt T-tau} model atmospheres. Overshooting was enabled in two convective regions: (i) above the hydrogen-burning core with $f=0.022$ and $f_0=0.002$; and (ii) at the stellar surface with $f=0.006$ and $f_0=0.001$ \citep{pedersenetal2018,pedersenetal2021}. We used a mixing length of $\alpha=1.9$. It appears that young, chemically homogeneous stars are insensitive to the specific value of $\alpha$ \citep{murphyetal2021a}.

We computed structure and pulsation models every 2\,Myr for ages between 50 and 180\,Myr. Our pulsation calculations included $\ell=0$ and $\ell=1$ modes, with frequencies up to 95\,d$^{-1}$. While so-called surface corrections are made for solar-like oscillators \citep[e.g.][]{kjeldsenetal2008,ball&gizon2014}, it seems that no such corrections are required for the predominantly radiative atmospheres of A stars \citep{beddingetal2020}, hence we made no such corrections here.

\subsection{Model Analysis}

The primary purpose of our models was to verify that our mode identifications agreed with models of reasonable mass, age, and metallicity. Nonetheless, it is useful to investigate how the non-rotating models perform at inferring stellar parameters. We investigated the three slowest rotators; the two rapid rotators were not modelled.

\begin{figure}
\begin{center}
\includegraphics[width=0.48\textwidth]{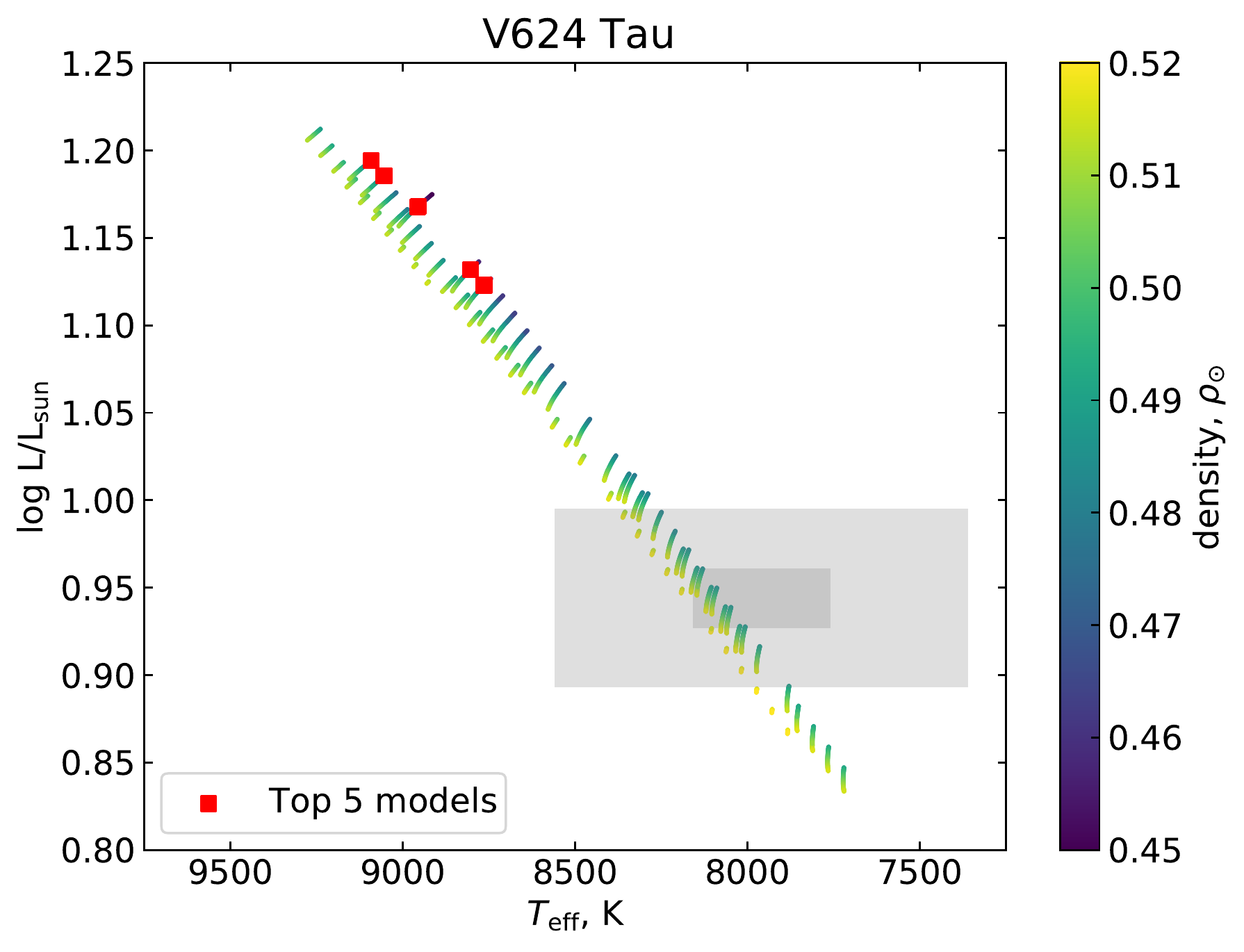}
\caption{H--R diagram for models of V624\,Tau. Dark and light grey boxes show the 1$\sigma$ and 3$\sigma$ bounds of the observed quantities. The best-fitting seismic models (red) are less dense than any model inside the 3$\sigma$ box.}
\label{fig:v624_hrd}
\end{center}
\end{figure}

Models were evaluated using a simple goodness-of-fit metric, based on summing the absolute values of the differences between the observed and model frequencies. The classical constraints, $T_{\rm eff}$ and $\log L$, were not used in model evaluation. For all three of the slowest rotators, the best-fitting models lie towards the most massive, most metal-rich, and oldest in the grid. In other words, the least dense models were consistently preferred. However, these models have temperatures and luminosities inconsistent with the observations at >5$\sigma$ (Fig.\,\ref{fig:v624_hrd}). We interpret this as an inadequacy of non-rotating models to accurately represent the oblate nature of rapidly rotating stars.

One way to create less dense non-rotating models is to change the initial helium abundance. We tested this with a 1.7\,M$_{\odot}$ star at [Fe/H] = 0.08 ($Z_{\rm init}$ = 0.0171). Reducing $Y_0$ by 1.8\%, from 0.280 to 0.275 (see eq.\,\ref{eq:dydz}), reduces the stellar density by 0.6\% at 50\,Myr and by 0.26\% at 180\,Myr. These lower-density models were a better fit to the observations than the models with higher helium content, keeping other parameters fixed.

Rotation, however, has a much larger effect. By creating otherwise identical models ($M = 1.7$\,M$_{\odot}$, $Z_{\rm init} = 0.0171$, $Y_0 = 0.280$) with a range of ZAMS rotation velocities of $v_{\rm in}$ = 30--270\,km\,s$^{-1}$, we noted that 
compared to a non-rotating model, the mean stellar density is $\sim$8\% lower for $v_{\rm in} = 150$\,km\,s$^{-1}$ and $\sim$37\% lower for $v_{\rm in} = 270$\,km\,s$^{-1}$. Rotation is therefore one to two orders of magnitude more important than the initial helium abundance, and the latter need only be varied when rotation is exceptionally well-constrained by asteroseismology.

In conclusion, rotating models will be required for accurate asteroseismic inference on the stellar parameters and these will be able to identify and use many more of the observed pulsation modes. Calculating such models and characterising these stars in the numerous extra model parameters this entails is beyond the scope of this work. A small sensitivity to the helium abundance suggests that this parameter, too, should be varied if rotation is well constrained.

\section{Conclusions}
\label{sec:conclusions}

We created custom light curves from K2 short-cadence data for five known $\delta$\,Sct stars in the Pleiades star cluster. We confirmed them to be high-frequency pulsators of the kind reported by \citet{beddingetal2020} and \citet{murphyetal2021a}. We identified several modes in each star and verified them using non-rotating models computed with {\sc mesa} and {\sc gyre}. We found very good agreement between theoretical and observational frequencies in models constructed around a narrow parameter space appropriate to the Pleiades cluster.
However, the lack of rotation in our models results in unrealistically high stellar densities that prevent accurate ages and metallicities from being derived. We therefore defer asteroseismic inference of these parameters for future work to include rotating models, which may also provide valuable constraints on the helium abundance in the Pleiades.

We infer the presence of rotationally split multiplets in our targets, which will allow the inference of stellar rotation rates. In particular, the dipole ridge of V647\,Tau comprises prograde dipole modes instead of the zonal modes that dominate the other stars. We also anticipate the arrival of three consecutive sectors of TESS data, which will increase the frequency resolution of the data, allowing more pulsation frequencies to be extracted. Unlike the K2 data, for which only five targets had the requisite short-cadence observations, dozens of targets will be available at 2-minute cadence in TESS, and the absence of thruster firings is expected to simplify the analysis.

\section*{Acknowledgements}

This research made use of {\sc Lightkurve}, a Python package for Kepler and TESS data analysis \citep{lightkurvecollaboration2018}, and \'Echelle \citep{hey&ball2020}. It also made use of the PASTEL catalogue, 2016 version \citep{soubiranetal2016}. We acknowledge support by the Australian Research Council through DP210103119 and FT210100485.
This work has made use of data from the European Space Agency (ESA) mission Gaia (\url{https://www.cosmos.esa.int/gaia}), processed by the Gaia Data Processing and Analysis Consortium (DPAC, \url{https://www.cosmos.esa.int/web/gaia/dpac/consortium}). Funding for the DPAC has been provided by national institutions, in particular the institutions participating in the Gaia Multilateral Agreement. This work also made use of ARI’s Gaia Services at \url{http://gaia.ari.uni-heidelberg.de/} for RUWE values.

\section*{Data Availability}

Jupyter notebooks detailing our creation of custom lightcurves from the K2 data are available at \url{https://github.com/hvidy/k2-pleaides-dsct}.


\bibliographystyle{mnras}
\interlinepenalty=10000
\bibliography{sjm_bibliography,trw_extra_refs,danhey_refs}
\bsp	



\label{lastpage}
\end{document}